# The rise of Indo-German collaborative research: 1990–2022

Aasif Ahmad Mir*[1], Nina Smirnova*[2], Jeyshankar Ramalingam[1], Philipp Mayr[2]

* Corresponding authors: miraasif7298@gmail.com; nina.smirnova@gesis.org

[1] Pondicherry University, Library and Information Science, Chinna Kalapet, Kalapet, Puducherry 605014, India.

[2] GESIS — Leibniz Institute for the Social Sciences, Unter Sachsenhausen 6-8, Cologne, 50667, Germany.

## Abstract

**Purpose:** The study aims to highlight the growth and development of Indo-German collaborative research over the past three decades. Moreover, this study encompasses an in-depth examination of funding acknowledgements to gain valuable insights into the financial support that underpins these collaborative endeavors. Together with this paper, we provide an openly accessible dataset of Indo-German research articles for further and reproducible research activities (the "Indo-German Literature Dataset").

**Methodology:** The data were retrieved from the Web of Science (WoS) database from the year 1990 till the 30th of November 2022. A total of 36,999 records were retrieved against the employed query. Acknowledged entities were extracted using a NER model specifically trained for this task. Interrelations between the extracted entities and scientific domains, lengths of acknowledgement texts, number of authors and affiliations, number of citations, and gender of the first author, as well as collaboration patterns between Indian and German funders were examined.

**Findings:** The study reveals a consistent and increasing growth in the publication trend over the years. The study brings to light that Physics, Chemistry, Materials Science, Astronomy & Astrophysics, and Engineering prominently dominate the Indo-German collaborative research. The United States, followed by England and France, are the most active collaborators in Indian and German research. Largely, research was funded by major German and Indian funding agencies, international corporations and German and American universities. Associations between the first author's gender and acknowledged entity were observed. Additionally, relations between entity, entity type, and scientific domain, were discovered.

**Implications:** The study paves the way for enhanced collaboration, optimised resource utilisation, and societal advantages by offering a profound comprehension of the intricacies inherent in research partnerships between India and Germany. Implementation of the insights gleaned from this study holds the promise of cultivating a more resilient and influential collaborative research ecosystem between the two nations.

**Originality:** The study highlights a deeper understanding of the composition of the Indo-German collaborative research landscape of the last 30 years and its significance in advancing scientific knowledge and fostering international partnerships. Furthermore, we provide an open version of the original WoS dataset. The Indo-German Literature Dataset consists of 22,844 articles from OpenAlex and is available for related studies like literature studies and Scientometrics.

# Introduction

The term 'research collaboration' is used to portray all forms of contract between academic institution's research, researchers, universities, R&D, and any combination of such two or more gatherings who share the obligation to reach a common goal by using their possessions available (Beaver and Rosen, 1995). The development of research collaboration will bring individual researchers together and support the sharing of knowledge between researchers, institutions, and countries. According to Sonnenwald (2007), scientists and organizations should consider the benefits and allotments of budgets for collaboration before deciding to collaborate. There have also been policies aimed at improving the links between science and technology through fostering research collaboration across sectors, in particular, between universities and industry.

Furthermore, most governments have been keen to increase the level of international collaboration engaged in by the researchers whom they support in the belief that this will bring about cost savings or other benefits. This study mainly concerned research collaboration between the Indian and German scientists who did research collaboratively and they were given acknowledgement by the funding agency for their research paper in Web of Science indexed journals.

The objectives of the following study are:

1. To study the year-wise growth pattern and citation trend of international collaborative research between India and Germany in the last 32 years.
    a. To highlight the subject-wise yearly growth of Indo-German scientific collaboration
2. To identify the top scientific disciplines, countries, and organizations, which are involved in Indo-German scientific collaboration.
3. To analyze acknowledgement texts of Indo-German research papers.

The first objective will address research questions 1 and 2:

RQ(1): What is the year-wise quantitative growth pattern of international collaborative research between India and Germany over the past 32 years?

RQ(2): How has the citation trend evolved for collaborative research between India and Germany during the specified period?

The second objective will focus on research questions 3,4,5 and 6:

RQ(3): What are the subject-wise yearly growth trends observed in Indo-German scientific collaborations?

RQ(4): Which scientific disciplines have shown prominent growth in collaborative research between India and Germany over the studied period?

RQ(5): Which are the top countries involved in collaborative research with India and Germany in various scientific disciplines over the last 32 years?

RQ(6): Which organizations have been actively engaged in Indo-German scientific collaboration?

The third objective will concentrate on research questions 7,8 and 9:

RQ(7): What are the general acknowledgement trends, i.e., which are the most acknowledged organizations, individuals, universities, or corporations?

RQ(8): Do highly-cited papers have more funding entities acknowledged, more authors and affiliations than papers with fewer citations?

RQ(9): What are general exchange patterns in acknowledgement texts, i.e. what are the most frequently collaborating German and Indian funding organizations?

## Related Work

### *Indo-German collaborative research*

India is Germany's most significant South Asian partner for scientific and technological cooperation, and they have established a strategic partnership in education, science, and research that serves as the foundation for future forms of collaboration (Khashimwo, 2015). To shed light on the history of collaboration between the two countries, Gupta and Fischer (2013) conducted a comparative analysis of science and technology (S&T) output in India and Germany from 2004 to 2009. Their study delved into various aspects, including the overall growth, impact, strengths and weaknesses of collaboration, geographical distribution, institutional participation, and collaborative linkages between institutions. Notably, the study revealed that Physical and Engineering sciences emerged as the stronger areas of Indo-German research collaboration. From the German side, Technical University Darmstadt, Ludwig Maximilians University Munich, and the University of Bonn stood out as the most active universities engaged in collaboration. On the Indian side, Panjab University in Chandigarh, Mangalore University, and the University of Delhi emerged as the most active collaborating institutions. These findings highlight the significant contribution of these universities in fostering collaborative research between India and Germany. In an earlier study by Rao and Gupta (2004), the research collaboration between India and Germany in the field of S&T was analyzed for the period of 1996 to 2000. The study unveiled a pattern of bilateral and multilateral collaboration evident in co-authored publications. Additionally, the research shed light on the involvement of important organizations, subjects of study, and the impact factor of collaborative research between the two countries.

Overall, these studies provide valuable insights into the collaborative research landscape between India and Germany, emphasizing their strong partnership in various scientific disciplines and laying the foundation for further advancements in science and technology.

### *Cross-Country studies*

Expanding beyond the Indo-German collaboration, cross-country studies reveal broader patterns of scientific collaboration. Gupta and Dhawan (2003) conducted a study on the collaborative research pattern between India and the People's Republic of China. They found evidence of growing research collaborations between the two countries. Interestingly, the study revealed that most of the science and technology (S&T) collaboration between China and India was conducted through multilateral channels, with limited progress through bilateral channels. In another study, Uddin and Singh (2014) presented a scientometrics analysis of academic research collaboration and output in the South Asian region. They assessed the level of research collaboration between South Asian nations and the rest of the world. Their findings showed that only a small percentage, around 2.2%, of papers involved collaboration among South Asian nations. On the other hand, a significant majority, approximately 97.7%, of collaborative output involved collaborations between South Asian and non-South Asian nations. Within the South Asian region, India emerged as the leading country in terms of research productivity and collaboration. Gupta et al. (2002) investigated scientific collaboration between India and Australia from 1995 to 1999. The study discovered that multilateral collaborative papers had a significantly higher average impact factor per paper compared to all papers. Additionally, approximately 38% of India-Australia collaborative papers involved other countries, including the United States, the United Kingdom, Russia, New Zealand, France, Switzerland, Italy, Japan, and China. Examining research collaboration between India and Latin America, Gupta and Singh (2004) analyzed S&T journals covered by SCI from 1996 to 2000. They identified Brazil, Mexico, Argentina, Columbia, Chile, Ecuador, Venezuela, and Peru as India's most significant collaborative partners in Latin America. The study revealed a strong multilateral partnership between India and 12 other Latin American nations, excluding Brazil and Mexico. Gupta and Dhawan (2006) explored India's position in science and technology, including its global share, ranking, strong and weak areas of research, quantity and quality of research output, and research dynamics across institutions, sectors, geographical regions, and subjects. Chakrabarti and Mondal (2021) investigated the research collaboration status between India and the five leading African countries over the past three decades. Their findings revealed that South Africa, Egypt, Ethiopia, Nigeria, and Kenya were India's top African collaborative partners. This highlights India's strong collaborative relationship with Northern African countries. Among developed countries, the United States, the United Kingdom, and Germany were identified as the most influential collaborating partners. Elango et al. (2021) conducted a quantitative and comparative analysis of India's and South Korea's scientific productivity. The study showed that South Korea had a higher proportion of publications with international collaboration compared to India. However, both nations continued to hold stronger positions in certain fields like chemical engineering and materials science.

Overall, these studies shed light on the collaborative research patterns and trends in various scientific domains involving India. They highlight the importance of international collaboration, the leading collaborating partners, and the need for increased research activities in specific fields. The existing literature on collaborative research between Germany and India, as well as other cross-country studies, offers valuable insights.

However, there are several limitations and gaps within the existing research landscape: Many studies confine their scope to specific time frames, such as intervals between 1995-2000 or 2004-2009. This narrow focus may fail to capture the dynamic shifts and recent trends in collaborative efforts. While certain research delves into disciplines like physical and engineering sciences, other fields that could potentially showcase substantial collaboration might be overlooked. This creates a partial picture of the overall collaborative landscape. Some studies tend to spotlight particular regions or institutions within the countries, inadvertently sidelining collaborative initiatives occurring elsewhere. This can overshadow the contributions of lesser-known entities. There's a dearth of exploration into the evolving nature of collaborative patterns over time and their potential transformations in the future. This limits our understanding of the adaptive nature of global partnerships. Existing literature often lacks in-depth policy prescriptions or strategic guidelines to enhance and foster collaborations between nations. There's a need for more actionable insights beyond outlining existing patterns. The focus on major nations or regions might lead to the neglect of collaborations involving smaller or less prominent countries, creating a skewed portrayal of global collaborative endeavours. Addressing these limitations holds the potential to significantly enhance our comprehension of collaborative research landscapes. It can facilitate more informed policy-making, enable better allocation of resources, and promote the development of stronger international partnerships.

### *Acknowledgements in scientific texts*

Acknowledgements in scientific articles as a rule express gratitude towards any help (financial, intellectual or technical) in conducting the research and writing an article. Information in the acknowledgement texts can be broadly distinguished into technical, instrumental and financial support or intellectual and conceptual support (Diaz-Faes and Bordons, 2017; Giles and Councill, 2004). Analysis of acknowledgements can reveal reward systems (Dziezyc and Kazienko, 2022), collaborative patterns and hidden research trends (Giles and Councill, 2004; Diaz-Faes and Bordons, 2017) in the scientific community. Acknowledged individuals can provide insights into informal research collaboration (Rose and Georg, 2021; Kusumegi and Sano, 2022). Acknowledged universities and corporations can show interaction and knowledge exchange between industry and universities (Chen et al., 2022). Project titles can reveal international scientific cooperation. Moreover, analysis of funding information can provide valuable insight into gender-specific differences in research funding, which are quite common in academia. Previous research revealed gender-specific differences in application behaviour (Ranga et al., 2012) and scoring of applications (Bol et al., 2022) along with less favourable assessments of women applicants (Witteman et al., 2019).

Most previous works of acknowledgements analysis used manually annotated data, and therefore were limited by the amount of processed data (Giles and Councill, 2004; Paul-Hus et al., 2017; Paul-Hus and Desrochers, 2019; Mccain, 2017) or relied on the information about funding agencies and grant numbers indexed in WoS or Scopus (Paul-Hus et al., 2016). Therefore, analysis was limited only to two types of acknowledged entities (funding organizations and grant numbers), disregarding other important information such as names of individuals, projects, universities or corporations. Moreover, the indexing of funding information on WoS and Scopus is incomplete and partly incorrect (Smirnova and Mayr, 2023a; Liu, 2020).

## Methodology & Data

The Web of Science (WoS), which is one of the most extensive indexing and abstracting databases, was consulted in December 2022 as a data source for the study. The following search query: CU ("GERMANY" AND "INDIA") was used to retrieve the data. The data were retrieved from the year 1990 till the 30th of November 2022. A total of 36,999 records were retrieved against the employed query.

Furthermore, we provide an open version of the original WoS dataset. The Indo-German Literature Dataset (Smirnova et al., 2024) consists of 22,844 articles from OpenAlex and is available for related studies like litterature studies and Scientometrics. OpenAlex (https://openalex.org/) serves as a freely accessible catalogue for the global research system, providing an open alternative to widely-used scientific knowledge bases such as Scopus and Web of Science (Priem et al., 2022; Culbert et al., 2024). The platform aggregates and standardizes data from various sources, including MAG, Crossref, and many others.[1] You can find a description and information on the process of the dataset's creation and its limitations in the dataset repository (Smirnova et al., 2024).

The WoS data were downloaded in text and MS Excel (xlsx) format for analysis. The text format data is imported to VOSviewer[2], to analyze the co-authorship / collaboration pattern of countries, organizations, and authors.

To analyze the "co-authorship" pattern of publications between countries, the "co-authorship" is set as a type of analysis and "countries" were set as a unit of analysis. Co-authorship analysis is performed based on the fractional counting method-means that the weight of a link is fractionalized e.g., if an author co-authors a document with 5 other authors, each of the 5 "co-authorship" links has a weight of 1/5. A total of 194 countries have participated in the Indo-German collaborative research. It was not possible to report each country in the study so in order to better visualize the data and provide insights about the highly productive countries we set a threshold of the minimum number of documents of a country: 1500. The set threshold was met by 22 out of 194 countries and thus 22 countries were analyzed for further analysis.

To analyze the "co-authorship" pattern of publications between organizations the "co-authorship" is set as a type of analysis and "Organizations" was set as a unit of analysis. To better visualize the data, we set a threshold of the minimum we set the threshold as minimum number of documents of an organization: 1000 (28 out of 39,960 organizations meet the threshold).

Additionally, an analysis of acknowledgement texts was conducted. Approximately half of the analyzed data had an acknowledgement text, which resulted in a corpus of 18,774 acknowledgement texts. Six types of acknowledged entities were extracted from the acknowledgement text: funding organizations (FUND), grant numbers (GRNB), universities (UNI), corporations (COR), persons (IND) and miscellaneous entities (MISC) (Figure 1).

---

[1] https://help.openalex.org/about-us
[2] https://www.vosviewer.com/

**IND** : person  
**FUND** : funding organization  
**GRNB** : grant number  
**UNI** : university  
**COR** : corporation  
**MISC** : miscellaneous

(1) Jan De Houwer is supported by Methusalem Grant BOF09/01M00209 of Ghent University and by the Interuniversity Attraction Poles Program initiated by the Belgian Science Policy Office (IUAPVII/33).

(2) Data on Anthem Blue Cross PPO enrollees were provided by Anthem, Inc.

*Figure 1: Examples of acknowledged entities. Different types of entities are highlighted with different colours.*

Acknowledged entities were extracted using a NER model specifically trained for this task (Smirnova and Mayr, 2023b). Extracted entities were additionally cleaned and disambiguated using the same procedure as described in Mayr and Smirnova (2023a). Interrelations between the extracted entities and scientific domains, lengths of acknowledgement texts, number of authors and affiliations, number of citations, and gender of the first author, as well as collaboration patterns between Indian and German funders were examined.

In the correlation analysis between gender and other variables, only the gender of the first author was considered. Gender was determined using the Python library gender-guesser[3]. Only the gender of authors whose full name was indexed in WoS could be predicted. Entries with androgynous names were excluded from the analysis. Only names which could be assigned to the category male or female were considered in the analysis. Names that fall into the category of mostly female were labelled as female and mostly male to male. The corpus for gender analysis was restricted to the criteria described above and resulted in a dataset of 5,359 entries (of a total of 18,774).

In the analysis of Indo-German cooperation, we searched for the patterns *'German'* or *'India'* in the name of the funding body and followed counted pairwise co-occurrences of funders in the same acknowledgement text.

---

[3] https://pypi.org/project/gender-guesser/

# Results

## *Yearly publication and citation trends of Indo-German collaborative research*

A total of 36,999 collaborative publications have been published by Indo-German researchers over the last three decades (1990-2022). By analyzing the yearly distribution of publications, it is evident that there has been increasing growth in the publication trend, and interestingly the trend is steadily rising every year. The majority of the documents were published during the recent seven years i.e., 2022 (2,840; 7.54%) followed by 2021 (3,206; 8.51%), 2020 (2,868; 7.62%), 2019 (2,587; 6.87%), 2018 (2,243; 5.95%), and 2017 (2,091; 5.55%) respectively. These seven years accounted for 47.43% of total collaborative publications. However, the minimum growth of the collaborative publications is observed during the starting phase of collaboration (1990-1997) during which only 4.7% of total publications were published.

On analyzing the yearly impact of publications in terms of citations, it is observed that publications published during the year 2014 (83,438) followed by 2013 (74,594), 2017 (73,599), 2015 (66,919), 2011 (63,922) and 2010 (62,208) respectively have received the highest number of citations. However, on analyzing the average citations per item (ACPP), it is evident that the publications published during the years 2003 (8,344.07), 2002 (5,581.24), 2009 (5,380.53), 2005 (5,212.54) 2004 (5,170.43) and 2008 (5,014.28) have received the highest score of citations per publication. Overall, a gradual increase with recurring ups and downs in the number of citations is observed over a period of time. Therefore, it can be inferred from the citation analysis of publications that collaboration can enhance the quality of scientific research and thereby improves the impact which is evident from the analysis that the publications have received 1,225,899 citations with an average of 32.57 citations per publication (Figure 2 & 3).

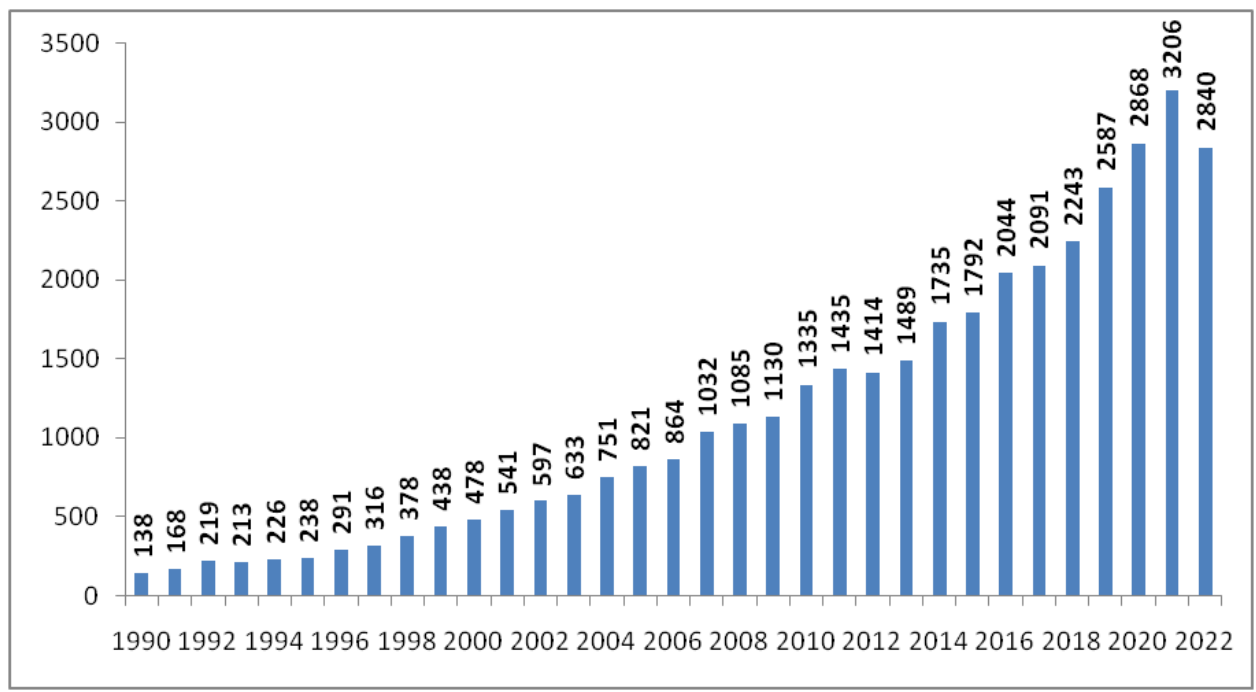

*Figure 2: Yearly Publication trends (The discrepancy observed between the total number of retrieved records and those depicted in the figure arises from publications where the year of publication is not explicitly stated)*

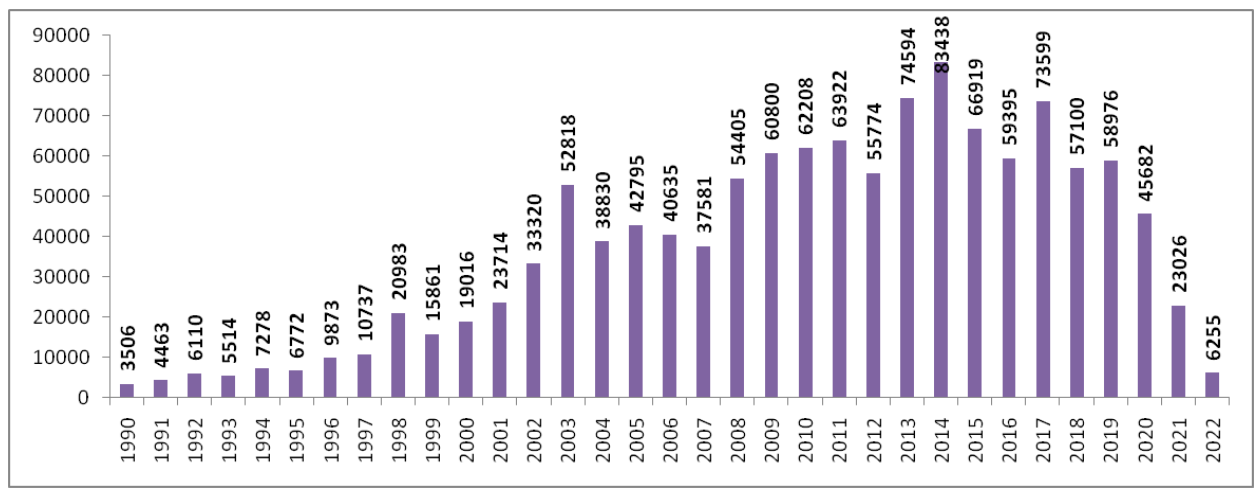

*Figure 3: Yearly Citation trends*

### *Top scientific disciplines of Indo-German scientific collaboration*

Due to the multidisciplinary nature of Indo-German collaborative research, the results demonstrate that a total of 151 disciplines are involved in the collaborative research between the two countries. However, for the analysis, only those disciplines were selected which contributed at least 500 documents. As depicted in Table 1, the results of quantitative analysis of the top scientific disciplines of Indo-German scientific collaboration highlight that Indo-German collaborative research is dominated by the Physics (9,565) and Chemistry (5,096) disciplines. However, a good number of publications were published by Materials Science (3,885) followed by Astronomy & Astrophysics (3,755), and Engineering (2,821) respectively. While the least priority collaborative research fields comprised of Microscopy (15), History & Philosophy of Science (15), Biomedical Social Sciences (14), Demography (14), Legal Medicine (13), Social Issues (12), Anatomy & Morphology (12), Cultural Studies (11), Women's Studies (11), Pathology (10), Architecture (10), Family Studies (9), Theater (8), Substance Abuse (6), Social Work (6), Art (5), Criminology & Penology (4), Nursing (2), Social Sciences - Other Topics (1), Medical Ethics (1) and Ethnic Studies (1).

*Table 1: Top 25 scientific disciplines of Indo-German scientific collaboration*

| Rank | Research Area | Record count |
| --- | --- | --- |
| 1 | Physics | 9,565 |
| 2 | Chemistry | 5,096 |
| 3 | Materials Science | 3,885 |
| 4 | Astronomy & Astrophysics | 3,755 |
| 5 | Engineering | 2,821 |
| 6 | Science & Technology - Other Topics | 2,226 |
| 7 | Environmental Sciences & Ecology | 1,557 |
| 8 | Computer Science | 1,509 |
| 9 | Biochemistry & Molecular Biology | 1,468 |
| 10 | Mathematics | 1,104 |
| 11 | Crystallography | 968 |
| 12 | Geology | 869 |
| 13 | Plant Sciences | 730 |
| 14 | Metallurgy & Metallurgical Engineering | 725 |

| 15 | Pharmacology & Pharmacy | 719 |
| --- | --- | --- |
| 16 | Neurosciences & Neurology | 706 |
| 17 | Oncology | 676 |
| 18 | Optics | 664 |
| 19 | Genetics & Heredity | 616 |
| 20 | Microbiology | 566 |
| 21 | Business & Economics | 549 |
| 22 | Polymer Science | 534 |
| 23 | Instruments & Instrumentation | 525 |
| 24 | Biotechnology & Applied Microbiology | 521 |
| 25 | Agriculture | 502 |

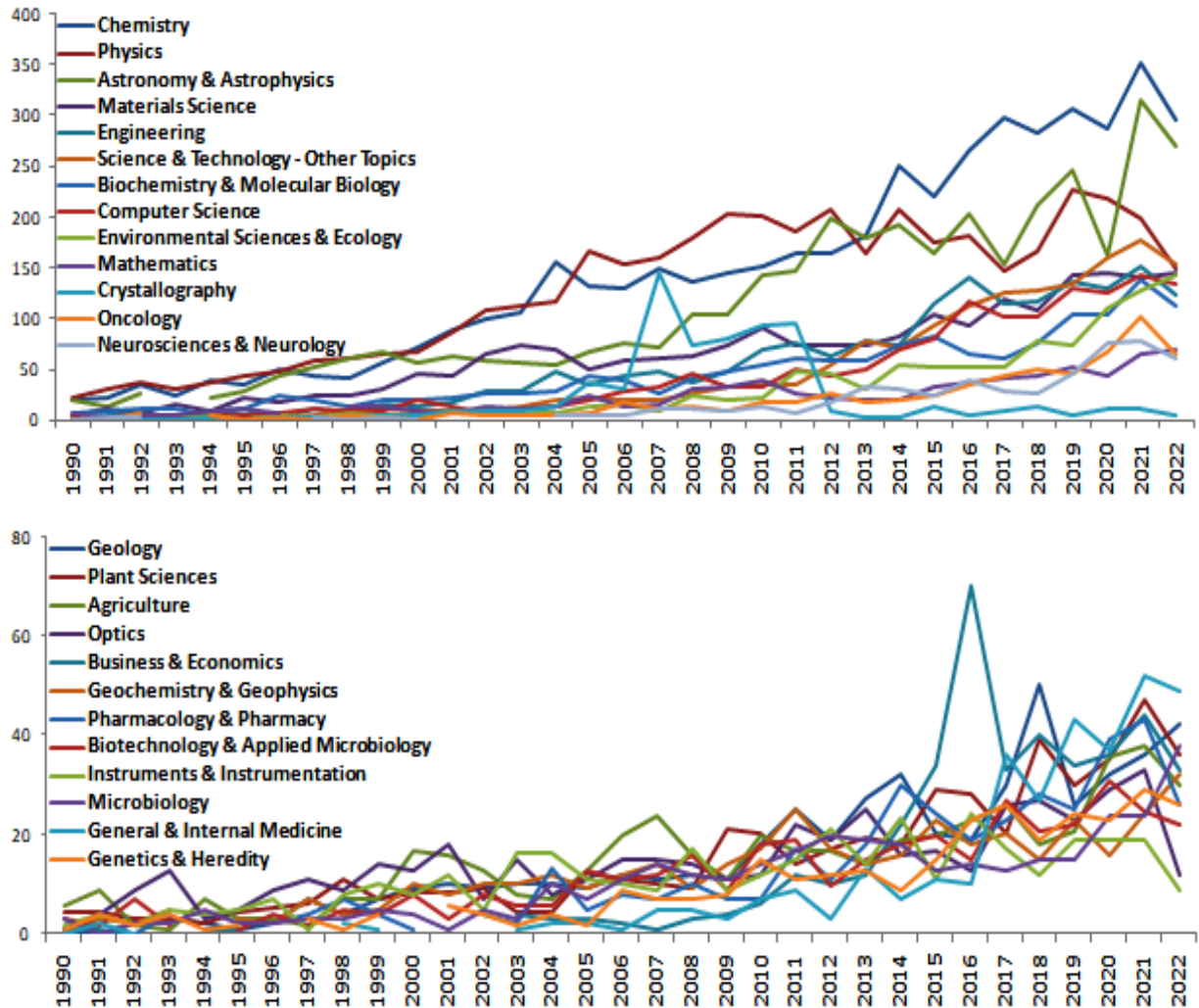

*Figure 4: Yearly disciplinary breakdown of the top 25 disciplines of Indo-German collaborative research from the year 1990-2022.*

## Co-authorship analysis between countries

Co-authorship is the most formal manifestation of intellectual collaboration in scientific research. It involves the participation of two or more authors in the production of a study, which leads to a scientific output of a greater quality or quantity than could be achieved by an individual (Hudson, 1996). A total of 194 countries have participated in the Indo-German collaborative research. However, only 22 countries meet the set threshold of 1500 documents. Other countries had less than 1,500 documents published and thus 22 countries were analyzed for country collaboration network analysis. As evident from Figure 5 there are two clusters (Cluster 1 Green: India and Germany, Cluster 2 Red: Other countries) linked together. These clusters help in revealing the author's country collaboration based on document weights. The analysis highlights that the USA (11,296 documents; with a TLS of 11,296) followed by England (6,127 documents; with a TLS of 6,127) and France (5,484 documents; with a TLS of 5,484) are the active collaborators in Indian and German research.

Therefore, the study revealed that there are two types of collaboration: one is bilateral in which only Indian and German institutions/scientists are involved and the other is multilateral in which scientists/institutions from countries like the USA, England, France, etc. also participated (Table 2).

*Table 2: Co-authorship analysis between countries*

| Country | Documents | Citations | TLS[4] |
|---|---|---|---|
| India | 36,310 | 1,148,719 | 35,660 |
| Germany | 36,180 | 1,143,742 | 35,623 |
| USA | 11,296 | 614,416 | 11,296 |
| England | 6,127 | 391,745 | 6,127 |
| France | 5,484 | 357,876 | 5,484 |
| Italy | 4,941 | 290,824 | 4,941 |
| China | 4,434 | 269,424 | 4,434 |
| Spain | 3,965 | 252,505 | 3,965 |
| Netherlands | 3,758 | 263,700 | 3,758 |
| Russia | 3,755 | 241,403 | 3,755 |
| Switzerland | 3,641 | 239,233 | 3,641 |
| Japan | 3,528 | 250,923 | 3,528 |
| Australia | 3,229 | 237,622 | 3,229 |

---

[4] TLS: The total link strength attribute indicates the total strength of the co-authorship links of a given researcher with other researchers. Total link strength attributes indicate, respectively, the number of links of an item with other items and the total strength of the links of an item with other items. For example, in the case of co-authorship links between researchers, the Links attribute indicates the number of co-authorship links of a given researcher with other researchers. The total link strength attribute indicates the total strength of the co-authorship links of a given researcher with other researchers.

| Canada | 3,145 | 276,259 | 3,145 |
| South Korea | 2,724 | 161,353 | 2,724 |
| Poland | 2,696 | 147,137 | 2,696 |
| Sweden | 2,525 | 194,024 | 2,525 |
| Brazil | 2,525 | 166,787 | 2,525 |
| Czech Republic | 1,975 | 97,469 | 1,975 |
| Austria | 1,785 | 101,033 | 1,785 |
| Belgium | 1,569 | 94,463 | 1,569 |
| Taiwan | 1,540 | 93,560 | 1,540 |

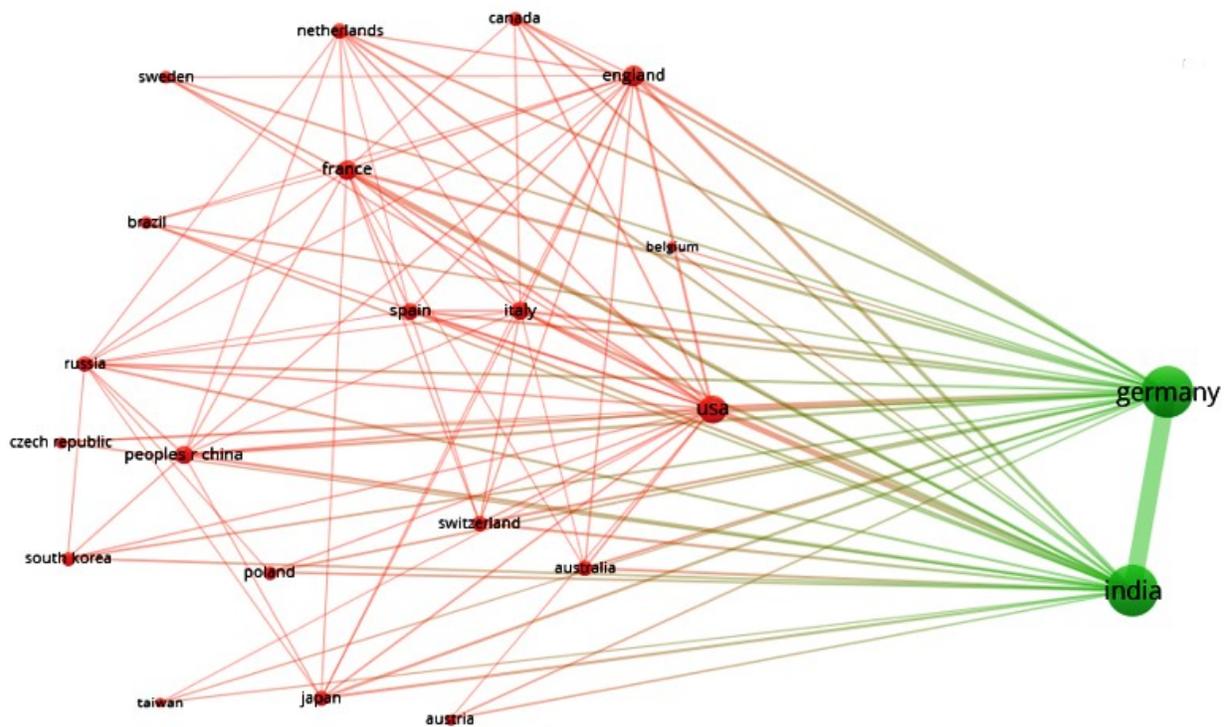

*Figure 5: Co-authorship network visualization between countries (network visualization based on document weights)*

## Co-authorship analysis between organizations

The co-authorship was set as the type of analysis and organizations were set as the unit of analysis. Threshold: Minimum number of documents of an organization: 1,000 (28 out of 39,960 organizations meet the threshold) for each of the 28 countries, the total link strength of the co-authorship links with other countries is calculated, and the countries with the greatest link strength are selected which are depicted in Table 3.

*Table 3: Co-authorship analysis between organizations*

| Organization | Documents | Citations | TLS |
| --- | --- | --- | --- |
| Indian Inst Technol | 3,848 | 126,853 | 1,064 |
| Tata Inst Fundamental Res | 2,252 | 102,628 | 1,420 |
| Inst High Energy Phys | 1,624 | 101,113 | 1,623 |
| Panjab Univ | 1,504 | 64,524 | 1,235 |
| Johannes Gutenberg Univ Mainz | 1,471 | 58,413 | 998 |
| Univ Sci & Technol China | 1,394 | 55,905 | 1,381 |
| Tech Univ Munich | 1,362 | 59,573 | 784 |
| Ist Nazl Fis Nucl | 1,359 | 97,180 | 1,294 |
| Indian Inst Sci | 1,347 | 36,747 | 362 |
| Univ Manchester | 1,333 | 82,435 | 1,177 |
| Rhein Westfal Th Aachen | 1,314 | 66,647 | 916 |
| Univ Illinois | 1,278 | 111,823 | 1,196 |
| Univ Michigan | 1,277 | 84,918 | 1,122 |
| Princeton Univ | 1,267 | 76,184 | 1,221 |
| Univ Bonn | 1,173 | 52,351 | 728 |
| Univ Calif Berkeley | 1,142 | 113,614 | 1,032 |
| Charles Univ Prague | 1,139 | 44,981 | 1,016 |
| Caltech | 1,136 | 113,742 | 1,053 |
| Univ Tokyo | 1,122 | 90,729 | 972 |
| Brookhaven Natl Lab | 1,113 | 109,753 | 1,094 |
| Bhabha Atom Res Ctr | 1,108 | 45,801 | 527 |
| Joint Inst Nucl Res | 1,099 | 45,668 | 1,081 |

| Indiana Univ | 1,099 | 43,422 | 1,057 |
| Heidelberg Univ | 1,063 | 69,164 | 560 |
| Chinese Acad Sci | 1,038 | 65,105 | 792 |
| Tech Univ Darmstadt | 1,020 | 13,336 | 344 |
| Korea Univ | 1,006 | 57,621 | 976 |
| Univ Delhi | 1,002 | 31,019 | 555 |

## *Cluster analysis of Indo-German research publications based on co-authorship*

The cluster analysis shows that there are three clusters with 28 items that are actively involved in Indo-German research collaboration. A graphic depiction of organizations involved in Indo-German research collaboration is displayed in Figure 6. The list of the most active organizations involved in Indo-German research collaboration can be found in Appendix D. The bigger the size of the node, the more active the organization is. Cluster 1 (red) contains ten organizations and is greatly dominated by the Institute of High Energy Physics China. Cluster 2 (green) comprised of nine organizations is greatly dominated by Tata Institute Fundamental Research India. Cluster 3 (blue), also composed of nine organizations is greatly dominated by the Indian Institute of Technology India.

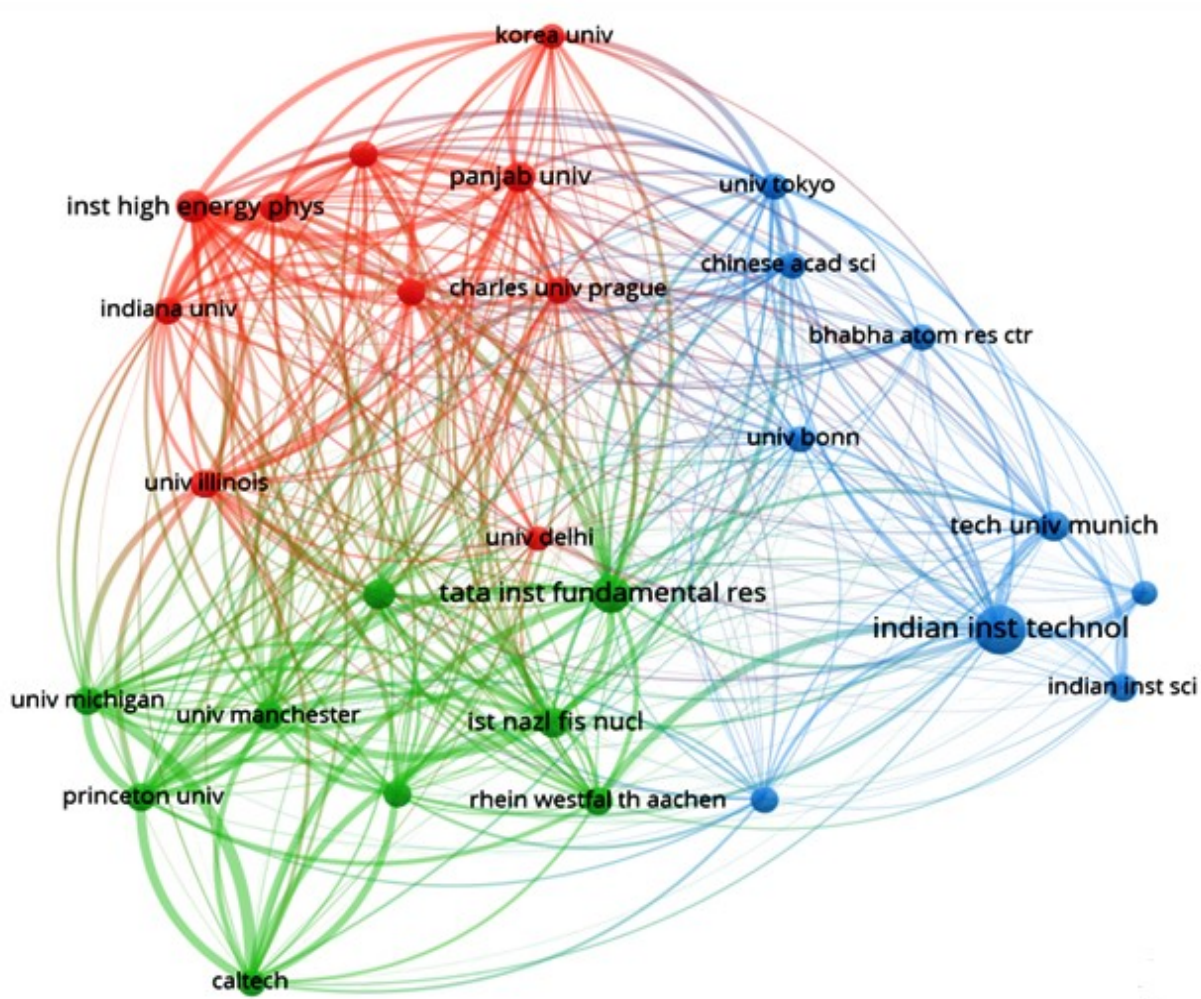

*Figure 6: Co-authorship analysis between organizations*

## *Acknowledgments analysis*

All analysis described in this section was performed using disambiguated data. Figure 7 shows the top 30 most acknowledged funding organizations (Figure 7-A), universities (Figure 7-B), corporations (Figure 7-C), and individuals (Figure 7-D). The German Research Foundation (DFG) and the Department of Science and Technology (DST) in India are the most acknowledged funders. Max Plank Society and Fermi National Accelerator Laboratory are the most acknowledged entities that fall into the university category. Novartis and Pfizer Inc. are the most acknowledged corporations. Overall, the distribution of funding organizations, corporations and universities follows a power law, also observed by Giles and Council (2004) and Smirnova and Mayr (2023a) in previous research. Thus, there is a small number of entities that have the greatest number of occurrences, while other entities occur massively less frequently. However, acknowledged individuals occur more evenly in the acknowledgements texts (see Figure 7-D).

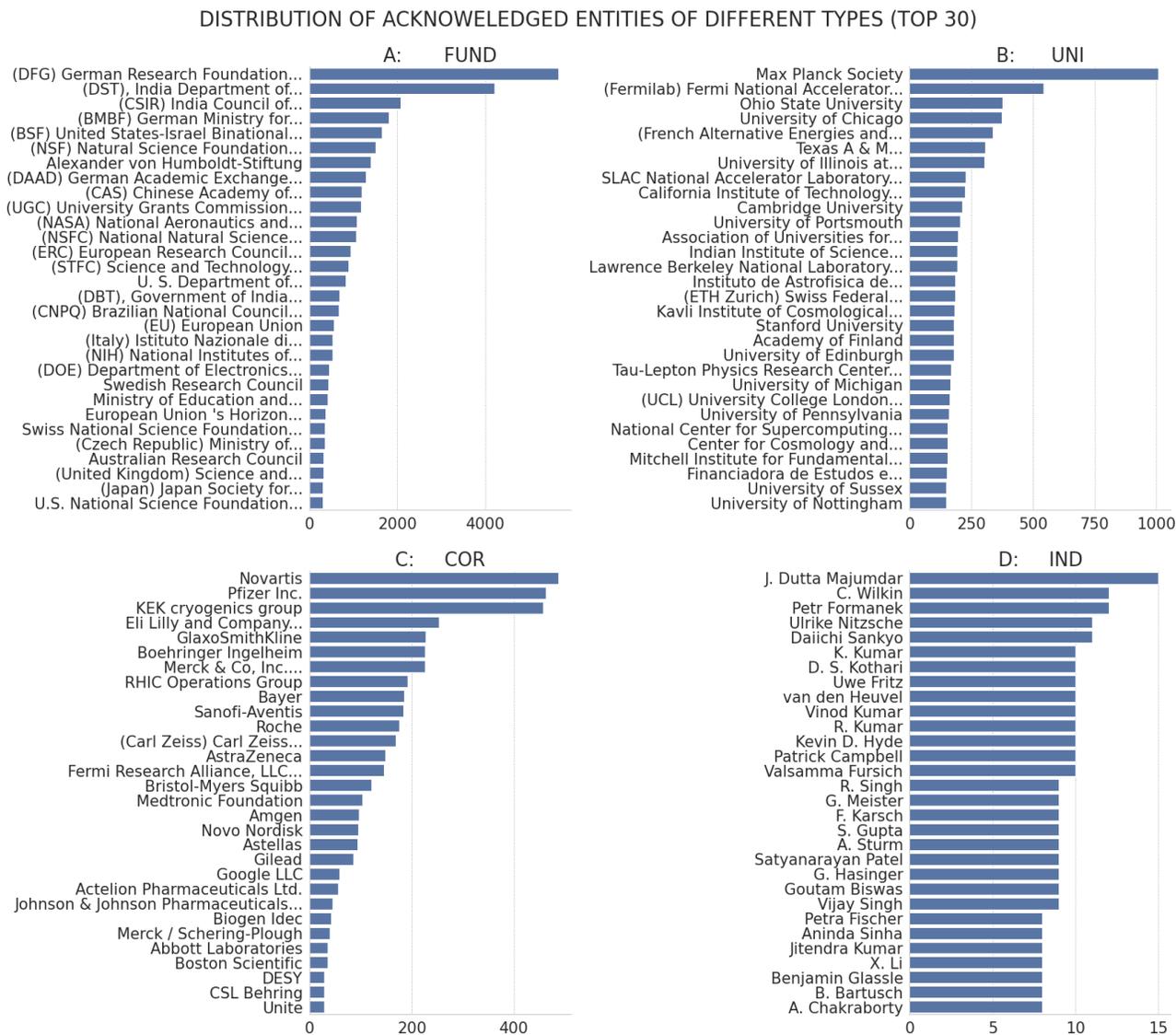

*Figure 7: Top 30 acknowledged entities of different types. Figure A represents the top 30 funding organizations, Figure B universities, Figure C corporations, and Figure D individuals.*

Quantitative variables were analyzed using the Pearson correlation (see Figure 8). A large correlation was found between the number of affiliations and the number of countries and authors. The number of funding organizations largely correlates with the number of countries, authors, affiliations and the total number of acknowledged entities in the acknowledgement text. A large correlation was also found between the number of funding organizations and grant numbers. Only a small correlation was observed between the number of citations, which the article obtains, and the number of countries, authors or affiliations, and acknowledged entities of all types. Which corresponds with the Smirnova and Mayr (2023a) findings.

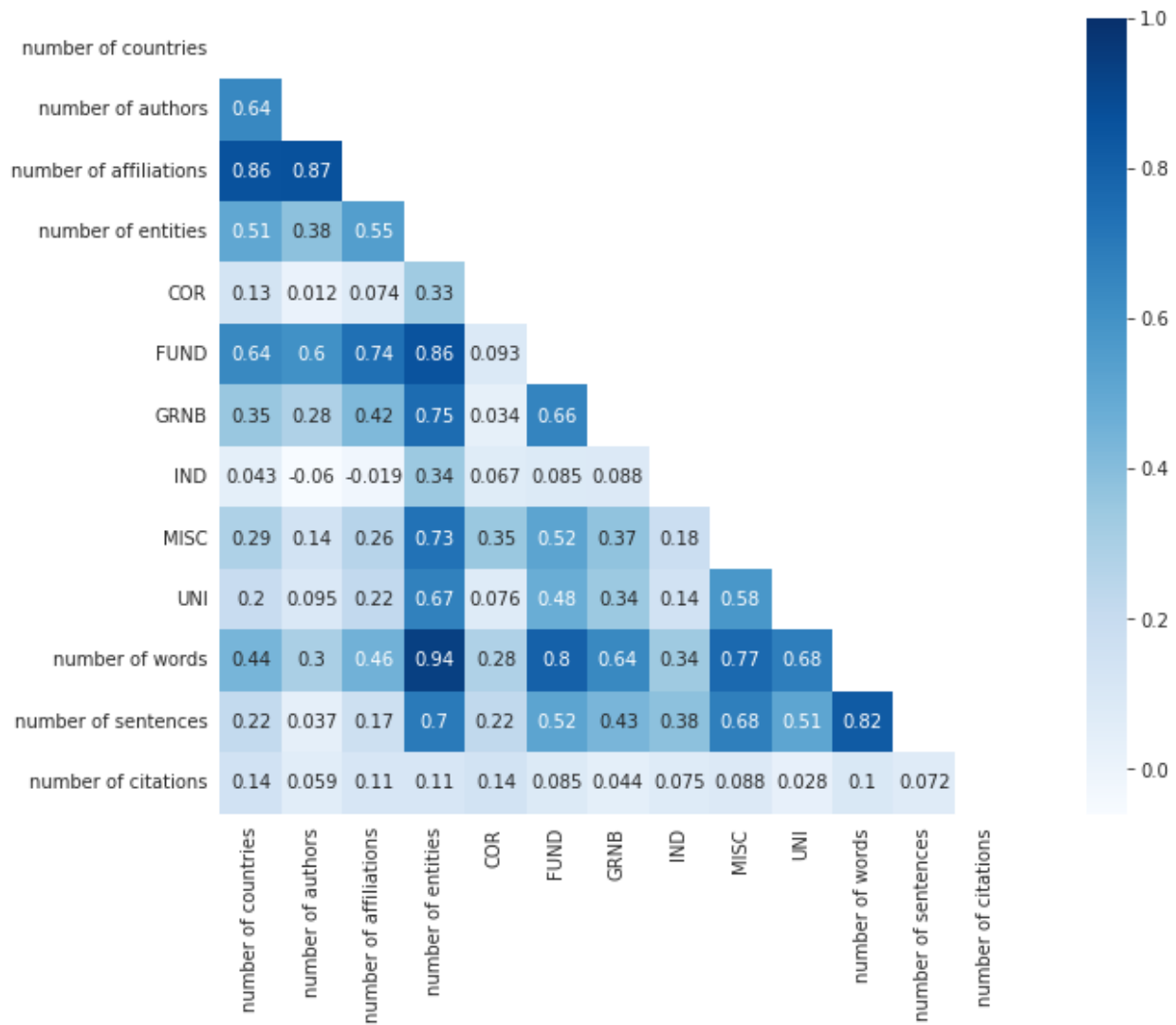

*Figure 8: Correlation between the number of citations of the article, number of acknowledged entities of different types, number of countries, number of authors, number of sentences and number of words in the acknowledgement text.*

The ANOVA test was performed on the analyzed variables. The P-value is very low (>0.005), which means that the results are statistically significant.

The strength of association between an acknowledged entity, entity type, and a scientific domain was assessed using Cramer's V test. The results of the analysis are shown in Figure 9. Table 4 demonstrates the degrees of freedom for the analyzed variable pairs. A strong association was observed between entity and entity type, scientific domain and entity, and gender and entity. Strong association indicates that a specific entity with a high possibility would belong to a specific entity type (which is expected), e.g., DFG is always a funding organization. At the same time, specific entities would occur more often or less often in the specific scientific domain and there is a connection between a first author's gender and an acknowledged entity. At the same time, only very weak associations were found between the first author's gender and entity type and scientific domain.

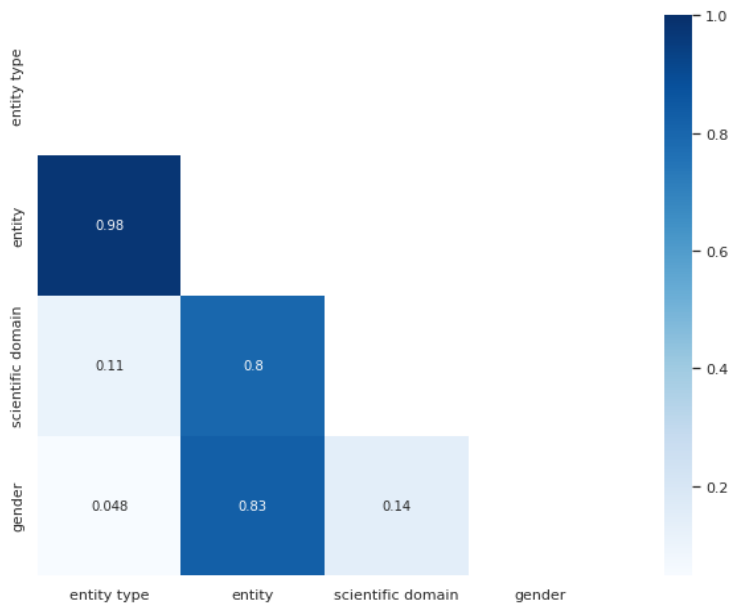

*Figure 9: Distribution of Cramer's V values between scientific domains, entity types, acknowledged entities and first author's gender.*

*Table 4: Degrees of freedom for the variable pairs*

|  | entity type | scientific domain |
|---|---|---|
| **entity type** |  | 5 |
| **entity** | 5 | 7 |
| **gender** | 1 | 1 |

As a strong association between entity and scientific domain and gender was found, we decided to look at different types of entities in detail. As Table 5 shows, strong associations were found between all entity types and scientific domains and the first author's gender.

Table 5: Cramer's V values and degrees of freedom between a specific entity type and first author's gender, and scientific domain

| Entity type | Cramer V | | Degrees of freedom | |
| --- | --- | --- | --- | --- |
| | scientific domain | gender | scientific domain | gender |
| FUND | 0.664306646 | 0.68791265 | 7 | 1 |
| GRNB | 0.915265227 | 0.70575796 | 7 | 1 |
| UNI | 0.717290889 | 0.7716894 | 6 | 1 |
| COR | 0.844310746 | 0.70575796 | 7 | 1 |
| IND | 0.958676355 | 0.97971835 | 6 | 1 |
| MISC | 0.85591922 | 0.90007502 | 7 | 1 |

Additionally, a Chi-square test of independence was conducted to examine whether there is a statistically significant relationship between entity, entity type, scientific domain and gender. As Table 6 demonstrates, the P-value for all variable pairs is less than 0.05, which indicates that the relation between these variables is significant.

Table: 6 The results of the Chi-square test for multiple variables. The column "Variables" represents variable pairs that were examined

| Variables | Chi-Square | P-value | Degrees of freedom |
| --- | --- | --- | --- |
| entity type-scientific domain | 14,843.56 | 0.00 | 35 |
| entity-scientific domain | 1,080,322 | 0.00 | 629,916 |
| entity-entity type | 1,157,013 | 0.00 | 450,095 |
| gender-entity | 37,952.55 | 9.46E-51 | 33,913 |
| gender-entity type | 126.0445 | 1.64E-25 | 5 |
| gender-scientific domain | 1,152.459 | 1.3E-244 | 7 |

Analysis of pairwise co-occurrences of Indian and German funders revealed the most frequent collaborators are the German Research Foundation (DFG), German Ministry for Education and Research (BMBF), German Academic Exchange Service (DAAD) and State of Niedersachsen from the German side, and Council of Scientific and Industrial

Research (CSIR), India, the Department of Science and Technology (DST), India and Department of Biotechnology (DBT), Government of India from the Indian side. Table 7 shows the top 10 frequent collaborating Indo-German funding organizations.

*Table 7: Top 10 Indo-German collaborating funding organizations*

| German funder | Indian funder | Frequency |
| --- | --- | --- |
| German Research Foundation (DFG) | Department of Science and Technology (DST), India | 1,217 |
| German Ministry for Education and Research (BMBF) | Department of Science and Technology (DST), India | 606 |
| German Research Foundation (DFG) | Council of Scientific and Industrial Research (CSIR), India | 429 |
| German Academic Exchange Service (DAAD) | Department of Science and Technology (DST), India | 415 |
| German Academic Exchange Service (DAAD) | Council of Scientific and Industrial Research (CSIR), India | 145 |
| German Research Foundation (DFG) | Department of Biotechnology (DBT), Government of India | 144 |
| German Ministry for Education and Research (BMBF) | Council of Scientific and Industrial Research (CSIR), India | 73 |
| German Ministry for Education and Research (BMBF) | Department of Biotechnology (DBT), Government of India | 55 |
| State of Niedersachsen / Germany | Council of Scientific and Industrial Research (CSIR), India | 43 |
| German Academic Exchange Service (DAAD) | Department of Biotechnology (DBT), Government of India | 43 |

## Findings and Discussion

The analysis of the yearly distribution of publications found a consistent and significant growth in the publication trend. Notably, this trend has been steadily rising each year, indicating an increasing interest in Indo-German collaborative research (see Figure 2). A quantitative examination of the top scientific disciplines involved in Indo-German scientific collaboration brings to light that Physics, Chemistry, Materials Science, Astronomy & Astrophysics, and Engineering prominently dominate this collaborative research (see Table 1 and Figure 4). These findings underscore the significance of these fields in fostering cooperation between India and Germany. A noteworthy aspect of the analysis is the involvement of a significant number of countries in Indo-German collaborative research. A total of 194 countries have participated in this collaboration, highlighting its global reach and extensive network. The study reveals the active collaborators in Indo-German research. The United States (USA) emerges as the leading collaborator followed by England and France respectively. Italy and China are the next top-ranked collaborators (see Table 2). These countries demonstrate a strong commitment to scientific collaboration with India and Germany. Moreover, the study found two types of collaboration involved in Indo-German research: Firstly, bilateral collaborations solely involve institutions and scientists from India and Germany. This type of partnership is more dominant (see e.g., Table 2) and showcases the unique synergy and exchange of knowledge between the two countries. Secondly, multilateral collaborations extend beyond India and Germany, encompassing scientists and institutions from other nations as well (see Table 3 and Figure 6). This highlights the diverse and inclusive nature of the Indo-German collaborative research landscape.

Analysis of acknowledgements texts revealed interesting patterns. Thus, a strong association was observed between the first author's gender and acknowledged entities. Therefore, some entities are mostly associated with male authors or with female authors, e.g., male or female authors are mostly supported by certain funders, corporations, universities and individuals. At the same time, a strong association between the first author's gender and scientific domain as well as between the scientific domain and the entity was revealed, which suggests that there is an interrelationship between the author's gender and scientific domain, also observed by Koning et al. (2021) and Kozlowski et al. (2022). This is an interesting observation, as only a weak direct association between the first author's gender and scientific domain was observed during the analysis.

As expected, a strong association between entity type and entity was found. The number of countries strongly correlates with the number of affiliations and funding organisations, which implies strong international collaboration patterns in the analysed research articles. The number of miscellaneous entities mildly correlates with the number of universities and funding organisations, which might indicate, that miscellaneous information mostly refers to these entity types, e.g., the name of the project associated with the funding organization or universities. Major German and Indian funding organizations were observed in the top 30 extracted funding organizations. International corporations were mostly represented in the top 30 corporations (see Figure 7). Universities were mostly represented by German and American research facilities. Strong collaboration patterns were mainly found between major German and Indian funders, such as DFG, BMBF and DAAD or DST, CSIR and DBT (see Table 7). Only a weak correlation between the number of citations and the number of affiliations, countries and entities was found. Analysis of extracted individuals provides an opportunity to investigate informal scientific collaboration. Nevertheless, names of individuals require additional disambiguation procedures, which poses further interesting challenges.

In the present paper, we conducted an initial gender analysis of funding behavior and examined general acknowledgement trends. We believe that acknowledgement data has a greater potential for further research, such as deeper gender analysis, exploration of differences between research domains, and expanding the analysis to other types of publications. Along with this, building upon the work of Santos et al. (2022), there's interest in delving deeper into the type of research, e.g. whether the funding was allocated for social or technology-based research.

Furthermore, we provided an open version (the Indo-German Literature Dataset[5]) of the data we used in our research using data from OpenAlex, which comprises 22,844 articles (Smirnova et al., 2024). Open OpenAlex data were gathered according to the doi for entries possessing a doi. For articles without doi, we conducted a matching procedure according to the article title, publishing year and authors. We believe that an open dataset on Indo-German publications will facilitate more research on this topic.

## Conclusions

The study shows an increasing interest and commitment to scientific cooperation between India and Germany. An interesting finding of this study is the extensive global reach of Indo-German collaborative research. This highlights the broad network and wide-ranging impact of this collaboration on a global scale. The study provides valuable insights into the fields of research driving this cooperation, the leading collaborators involved, and the collaborative models employed. These findings contribute to a deeper understanding of the Indo-German collaborative research landscape and highlight its significance in advancing scientific knowledge and fostering international partnerships.

---

[5] https://doi.org/10.5281/ZENODO.10607234

Analysis of the acknowledgement texts revealed the most frequent funders, universities and corporations. Largely, research was funded by major German and Indian funding agencies, international corporations and German and American universities. The most frequent Indo-German collaboration relations were also observed between major German and Indian funders. Associations between the first author's gender and acknowledged entity were observed. Additionally, relations between entity, entity type, and scientific domain, were discovered. Only a very weak correlation was found between the number of citations and the number of authors, affiliations, countries and funding organisations. Overall, the findings of this study provide unique insights into the collaborative research dynamics between India and Germany, opening the way for improved scientific cooperation, resource optimisation, and the possibility for significant social benefits.

## Acknowledgement

This work was funded by DAAD–UGC Project-based Personnel Exchange Programme (PPP 2022) via the project "Social Network and Scientometric Analysis in Collaborative Research Publications between India and Germany" (SEASON -- https://github.com/kalawinka/season/), grant numbers: DAAD project number: 57608852; UGC project number: 1-10/2020(IC). Nina Smirnova received funding from the German Research Foundation (DFG) via the project "POLLUX" under grant number MA 3964/7-2.

# References

Anuradha, K. and Urs, S. (2007), "Bibliometric indicators of Indian research collaboration patterns: A correspondence analysis", *Scientometrics,* Vol. 71 No. 2, pp.179-189.

Bala, A. and Gupta, B. M. (2010), "Mapping of Indian neuroscience research: A scientometric analysis of research output during 1999-2008", *Neurology India,* Vol. 58 No.1, pp. 35.

Beaver, D. and R. Rosen (1978), "Studies in Scientific Collaboration: Part I — The Professional Origins of Scientific Co-authorship", *Scientometrics,* Vol.1 No.1, pp. 65-84.

Bol, T. De Vaan, M. and Van De Rijt, A. (2022), "Gender-equal funding rates conceal unequal evaluations", *Research Policy*, Vol. 51 No.1, https://doi.org/10.1016/j.respol.2021.104399

Chakrabarti, K. and Mondal, D. (2021), "India's Research Collaboration Trend with the Selected African Countries: An Exploratory Study", *Journal of Scientometric Research,* Vol.10 No.3, pp. 412-422.

Chaurasia, N. K. Chavan, S. B. and Kataria, S. (2018), "Mapping the Research Collaboration Between India And USA During 2007–2016. *In 2018 5th International Symposium on Emerging Trends and Technologies in Libraries and Information Services (ETTLIS)",* pp. 338-343.

Chen, H. Song, X. Jin, Q. and Wang, X. (2022). "Network dynamics in university-industry collaboration: A collaboration-knowledge dual-layer network perspective", *Scientometrics,* Vol.127 No.11, pp. 6637–6660. https://doi.org/10.1007/s11192-022-04330-9

Culbert, J., Hobert, A., Jahn, N., Haupka, N., Schmidt, M., Donner, P. and Mayr, P. (2024). "Reference Coverage Analysis of OpenAlex compared to Web of Science and Scopus", http://arxiv.org/abs/2401.16359

Hemalatha, M. Rajendran, M. and Rao, C. (2022). "Mapping the research impact of collaboration and networking of ICAR fisheries research institutes in India: A scientometric study", *Indian Journal of Fisheries,* Vol.69 No.1, pp. 1-21.

Díaz-Faes, A. and Bordons, M. (2017). "Making visible the invisible through the analysis of acknowledgements in the humanities". *Aslib Journal of Information Management,* Vol.*69* No.5, pp. 576–590. https://doi.org/10.1108/AJIM-01-2017-0008

Dzieżyc, M. and Kazienko, P. (2022). "Effectiveness of research grants funded by European Research Council and Polish National Science Centre", *Journal of Informetrics,* Vol. 16 No.1, pp. 101243.https://doi.org/10.1016/j.joi.2021.101243

Elango, B. Oh, D. G. and Rajendran, P. (2021). "Assessment of scientific productivity by India and South Korea", *DESIDOC Journal of Library and Information Technology,* Vol.41 No.3, pp.190-198.

Elango, B. Rajendran, P. and Bornmann, L. (2013). "Global nanotribology research output (1996–2010): A scientometric analysis", *Plos One,* Vol.8 No.12.

Garg, K. C. and Kumar, S. (2018). "Collaboration patterns of Indian scientists in organic chemistry", *Current Science,* Vol.114 No.6, pp.1174-1180. https://www.jstor.org/stable/26797322

Giles, C. L. and Councill, I. G. (2004). "Who gets acknowledged: Measuring scientific contributions through automatic acknowledgment indexing", *Proceedings of the National Academy of Sciences,* Vol.101 No.51, pp.17599–17604. https://doi.org/10.1073/pnas.0407743101

Gupta, B. M. and Bala, A. (2011). "A scientometric analysis of Indian research output in medicine during 1999–2008", *Journal of natural science, biology, and medicine,* Vol.2 No.1, pp.87.

Gupta, B. M. and Dhawan, S. M. (2003). "India's Collaboration with People's Republic of China in Science and Technology: A Scientometric Analysis of Coauthored Papers During 1996-2000", *China Report,* Vol.39 No.2, pp. 197-211.

Gupta, B. M. and Fischer, T. (2013). "Indo-German collaborative research during 2004-09: An quantitative assessment", *Indian Journal of Science and Technology*, Vol.6 No.2, pp. 177-191. https://doi.org/10.17485/ijst/2013/v6i2.24

Gupta, B. M. and Singh, M. (2004). "India's collaboration with Latin America as reflected in co-authored papers", *DESIDOC Journal of Library and Information Technology*, Vol.24 No.3.

Gupta, B. M. Dhawan, S. M. Bose, P. R. and Mishra, P. K. (2002). "India's collaboration with Australia in science and technology: A Scientometric study of co-authored papers during 1995-199", *DESIDOC Journal of Library and InformationTechnology,* Vol.22 No.6.

Gupta, B. M. Munshi, U. M. and Mishra, P. K. (2004). "Regional collaboration in SandT among South Asian countries", *Annals of Library and Information Studies,* Vol. 51 No.4, pp. 121-132

Gupta, B. M. Singh, N. and Gupta, R. (2015). "Indian cloud computing research: A scientometric assessment of publications output during 2004-1", *SRELS Journal of Information Management,* Vol.52 No.5, pp.315-326.

Gupta, B.M. and Dhawan, S.M. (2006). "Measures of Progress of Science in India: An Analysis of the Publication Output in Science and Technology", *New Delhi; Office of the Principal Scientific Adviser to the Government of India.* (http://psa.gov.in/writeraddata/11913286541_MPSI.pdf)

Hudson, J. (1996). "Trends in Multi-Authored Papers in Economics", *Journal of Economic Perspectives,* Vol.10 No.3, pp. 153–158. https://doi.org/10.1257/jep.10.3.153

Jhamb, G. Meera, and Singh, K. P. (2019). "Indian geology research as reflected by Web of Science during 1998-2017", *COLLNET Journal of Scientometrics and Information Management,* Vol.13 No.1, pp. 37-51.

Kappi, M. and Biradar, B. S. A. (2022). "Comparative Bibliometric Analysis and Visualization of Indian and South Korean Library and Information Science Research Publications During 2001–2020", *International Journal of Knowledge Content Development and Technology*.

Kaur, H. and Mahajan, P. (2015). "Collaboration in medical research: a case study of India", *Scientometrics,* Vol.105 No.1, pp. 683-690. https://doi.org/10.1007/s11192-015-1691-6

Khashimwo, P. (2015). "India and Germany: Global Partnership in 21st century", *International Journal of scientific research and management (IJSRM),* Vol.3 No 6, pp. 3188-3195.

Koning, R. Samila, S. and Ferguson, J.-P. (2021). "Who do we invent for? Patents by women focus more on women's health, but few women get to invent", *Science,* Vol.372 No.6548, pp. 1345–1348. https://doi.org/10.1126/science.aba6990

Kousalya, K. and Jeyshankar, R. (2021). "Collaborative Research Publications Trends on Disaster Management: A Scientometric Analysis", *Library Philosophy and Practice*, pp.1-23.

Kozlowski, D. Larivière, V. Sugimoto, C. R. and Monroe-White, T. (2022). "Intersectional inequalities in science", *Proceedings of the National Academy of Sciences,* Vol.119 No.2,.https://doi.org/10.1073/pnas.2113067119

Kumar, A. Mallick, S. and Swarnakar, P. (2020). "Mapping Scientific Collaboration: A Bibliometric Study of Rice Crop Research in India", *Journal of Scientometric Research.*

Kumar, S. and K. C. Garg. (2005). "Scientometrics of computer science research in India and China", *Scientometrics*, Vol. 64 No.2, pp.121–32. doi:10.1007/s11192-005-0244-9.

Kusumegi, K. and Sano, Y. (2022). "Dataset of identified scholars mentioned in acknowledgement statements", *Scientific Data,* Vol.9 No.1, pp.461. https://doi.org/10.1038/s41597-022-01585-y

Liu, W. (2020). "Accuracy of funding information in Scopus: A comparative case study", *Scientometrics,* Vol.124 No.1, pp. 803–811. https://doi.org/10.1007/s11192-020-03458-w

McCain, K. W. (2018). "Beyond Garfield's Citation Index: An assessment of some issues in building a personal name Acknowledgments Index", *Scientometrics,* Vol.114 No.2, pp. 605–631.https://doi.org/10.1007/s11192-017-2598-1

Mohan, S. Gupta, B. M. and Dhawan, S. M. (2003). "Materials science research and development in India: A scientometric analysis of international collaborative output", *DESIDOC Journal of Library and Information Technology,* Vol.23 No.2.

Pathak, M. (2020). "Quantitative analysis of international collaboration on COVID-19: Indian perspective", *Indian Journal of Biochemistry and Biophysics (IJBB),* Vol. 57 No.4, pp. 439-443.

Pathak, M. and Kumari, N. K. (2019). "Pharmaceutical Research in India: A Scientometric Analysis of International Collaboration", *JSIR,* Vol.78 No.11, http://nopr.niscpr.res.in/handle/123456789/51198

Paul-Hus, A. and Desrochers, N. (2019). "Acknowledgements are not just thank you notes: A qualitative analysis of acknowledgements content in scientific articles and reviews published in 2015" *PLOS ONE,* Vol.14 No.12, pp. e0226727. https://doi.org/10.1371/journal.pone.0226727

Paul-Hus, A. Desrochers, N. and Costas, R. (2016). "Characterization,description, and considerations for the use of funding acknowledgement data in Web of Science", *Scientometrics,* Vol.108 No,), pp.167–182. https://doi.org/10.1007/s11192-016-1953-y

Paul-Hus, A. Díaz-Faes, A. A. Sainte-Marie, M. Desrochers, N. Costas, R. and Larivière, V. (2017). "Beyond funding: Acknowledgement patterns in biomedical, natural and social sciences", *PLOS ONE,* Vol.12 No.10. https://doi.org/10.1371/journal.pone.0185578

Priem, J., Piwowar, H. and Orr, R. (2022). "OpenAlex: A fully-open index of scholarly works, authors, venues, institutions, and concepts", http://arxiv.org/abs/2205.01833

Ranga, M. Gupta, N. and Etzkowitz, H. (2012). "*Gender Effects in Research Funding",* https://www.dfg.de/download/pdf/dfg_im_profil/geschaeftsstelle/publikationen/studien/studie_gender_effects.pdf

Rao, M.K.D. and Gupta, B.M. (2004), "Indo-German collaboration in SandT: An analysis through coauthored publications", *Annals of Library and Information Studies,* Vol.51 No.2,pp. 64-71.

Rose, M. E. and Georg, C.-P. (2021). "What 5,000 acknowledgements tell us about informal collaboration in financial economics", *Research Policy,* Vol.50 No.6, pp.104236. https://doi.org/10.1016/j.respol.2021.104236

Santos, J.M., Horta, H. & Luna, H. The relationship between academics' strategic research agendas and their preferences for basic research, applied research, or experimental development. *Scientometrics* **127**, 4191–4225 (2022). https://doi.org/10.1007/s11192-022-04431-5

Shrivastava, R. and Mahajan, P. (2016). "Artificial intelligence research in India: a scientometric analysis", *Science and Technology Libraries,* Vol.35 No.2, pp.136-151.https://doi.org/10.1080/0194262X.2016.1181023

Smirnova, N., Culbert, J. H. and Mayr, P. (2024). "Indo-German Literature Dataset" [dataset], Zenodo. https://doi.org/10.5281/ZENODO.10607234

Smirnova, N. and Mayr, P. (2023a). "A comprehensive analysis of acknowledgement texts in Web of Science: A case study on four scientific domains", *Scientometrics,* Vol.128 No.1, pp.709–734. https://doi.org/10.1007/s11192-022-04554-9

Smirnova, N. and Mayr, P. (2023b). "Embedding Models for Supervised Automatic Extraction and Classification of Named Entities in Scientific Acknowledgements", *Scientometrics* (2023). https://doi.org/10.1007/s11192-023-04806-2

Sonnenwald, D. H. (2007). "Scientific collaboration", *Annual Review of Information Science and Technology,* Vol.41 No.1, pp. 643-681.

Uddin, A. and Singh, V. K. (2014). "Measuring research output and collaboration in South Asian countries", *Current Science,* pp.31-38.

Wainer, J. E. C. Xavier, and F. Bezerra. (2009). "Scientific production in computer science: A comparative study of Brazil and other countries", *Scientometrics,* Vol.81 No.2, pp. 535–47.doi:10.1007/s11192-008-2156-y.

Witteman, H. O. Hendricks, M. Straus, S. and Tannenbaum, C. (2019). "Are gender gaps due to evaluations of the applicant or the science? A natural experiment at a national funding agency", *The Lancet*, Vol.393 No.10171, pp. 531–540. https://doi.org/10.1016/S0140-6736(18)32611-4

# Appendix A

## Yearly disciplinary breakdown of Indo-German collaborative research from 1990-2000

| Research area | 1990 | 1991 | 1992 | 1993 | 1994 | 1995 | 1996 | 1997 | 1998 | 1999 | 2000 |
|---|---|---|---|---|---|---|---|---|---|---|---|
| Chemistry | 21 | 23 | 37 | 26 | 40 | 36 | 52 | 45 | 43 | 57 | 73 |
| Physics | 22 | 31 | 38 | 31 | 37 | 43 | 49 | 58 | 62 | 65 | 67 |
| Astronomy & Astrophysics | 21 | 14 | 26 |  | 22 | 29 | 45 | 53 | 61 | 67 | 58 |
| Materials Science | 6 | 13 | 8 | 17 | 10 | 24 | 19 | 25 | 26 | 31 | 47 |
| Engineering | 3 | 6 | 7 | 3 | 9 | 12 | 4 | 11 | 7 | 16 | 14 |
| Science & Technology - Other Topics | 4 | 3 | 2 | 1 | 5 | 3 | 4 | 5 | 11 | 11 | 9 |
| Biochemistry & Molecular Biology | 7 | 9 | 12 | 12 | 5 | 12 | 24 | 21 | 13 | 21 | 21 |
| Computer Science | 2 | 2 | 4 | 2 | 5 | 5 | 6 | 10 | 8 | 8 | 20 |
| Environmental Sciences & Ecology | 0 | 2 | 4 | 2 | 4 | 1 | 4 | 2 | 7 | 5 | 5 |
| Mathematics | 7 | 2 | 5 | 4 | 9 | 11 | 6 |  | 12 | 12 | 8 |
| Crystallography | 0 | 1 | 0 | 1 | 2 | 1 | 3 | 3 | 3 | 5 | 5 |
| Oncology | 2 | 1 | 7 |  | 5 | 2 | 3 | 1 | 5 | 4 | 1 |
| Neurosciences & Neurology | 0 | 2 | 3 |  | 2 |  |  | 2 | 1 | 2 | 1 |
| Geology | 1 | 0 | 2 | 3 | 2 | 1 | 2 | 7 | 4 | 7 | 9 |
| Plant Sciences | 4 | 4 | 3 | 3 | 2 | 4 | 5 | 6 | 11 | 7 | 8 |
| Agriculture | 6 | 9 | 2 | 1 | 7 | 3 | 3 | 2 | 7 | 7 | 17 |
| Optics | 1 | 4 | 9 | 13 | 3 | 5 | 9 | 11 | 9 | 14 | 13 |
| Business & Economics | 1 | 1 | 0 |  | 1 | 1 |  | 1 |  | 1 |  |
| Geochemistry & Geophysics | 0 | 3 | 2 | 5 |  | 3 | 2 | 7 | 4 | 5 | 10 |
| Pharmacology & Pharmacy | 3 | 0 | 0 | 4 |  |  | 2 | 4 | 7 | 4 | 1 |
| Biotechnology & Applied Microbiology | 3 | 2 | 7 | 2 | | 1 | 4 | 2 | 5 | 4 | 8 |
| Instruments & Instrumentation | 2 | 4 | 2 | 5 | 4 | 5 | 7 | 1 | 8 | 10 | 8 |
| Microbiology | 3 | 0 | 2 | 2 | 5 | 2 | 2 | 3 | 3 | 5 | 4 |
| General & Internal Medicine | 0 | 2 | 0 | | 2 | | | | 2 | 1 | |

# Appendix B

**Yearly disciplinary breakdown of Indo-German collaborative research from 2001-2011**

| Research area | 2001 | 2002 | 2003 | 2004 | 2005 | 2006 | 2007 | 2008 | 2009 | 2010 | 2011 |
|---|---|---|---|---|---|---|---|---|---|---|---|
| Chemistry | 89 | 100 | 106 | 157 | 133 | 131 | 149 | 137 | 146 | 153 | 166 |
| Physics | 86 | 109 | 114 | 117 | 167 | 155 | 161 | 180 | 204 | 201 | 187 |
| Astronomy Astrophysics | 64 | 60 | 57 | 55 | 68 | 76 | 73 | 104 | 104 | 144 | 148 |
| Materials Science | 44 | 65 | 74 | 71 | 51 | 60 | 62 | 63 | 75 | 92 | 75 |
| Engineering | 18 | 28 | 28 | 49 | 35 | 43 | 49 | 38 | 49 | 70 | 75 |
| Science Technology - Other Topics | 10 | 9 | 13 | 21 | 22 | 21 | 21 | 26 | 33 | 36 | 36 |
| Biochemistry Molecular Biology | 23 | 27 | 27 | 29 | 43 | 39 | 26 | 41 | 48 | 54 | 62 |
| Computer Science | 14 | 6 | 11 | 14 | 20 | 28 | 32 | 45 | 33 | 32 | 50 |
| Environmental Sciences & Ecology | 8 | 8 | 11 | 6 | 14 | 15 | 8 | 24 | 20 | 21 | 47 |
| Mathematics | 6 | 14 | 11 | 14 | 25 | 14 | 15 | 31 | 33 | 40 | 27 |
| Crystallography | 8 | 8 | 8 | 13 | 36 | 30 | 146 | 74 | 81 | 94 | 96 |
| Oncology | 7 | 5 | 6 | 5 | 7 | 18 | 12 | 13 | 9 | 19 | 18 |
| Neurosciences Neurology | | 3 | | 4 | 5 | 4 | 11 | 10 | 9 | 13 | 6 |
| Geology | 10 | 9 | 10 | 10 | 9 | 12 | 11 | 12 | 11 | 18 | 25 |
| Plant Sciences | 8 | 9 | 4 | 4 | 12 | 11 | 10 | 9 | 21 | 20 | 14 |
| Agriculture | 16 | 13 | 8 | 7 | 13 | 20 | 24 | 16 | 11 | 20 | 17 |
| Optics | 18 | 7 | 15 | 8 | 11 | 15 | 15 | 14 | 11 | 12 | 22 |
| Business Economics | 1 | | 4 | 3 | 3 | 2 | 1 | 3 | 4 | 6 | 12 |

| | | | | | | | | | | |
|---|---|---|---|---|---|---|---|---|---|---|
| Geochemistry Geophysics | 8 | 10 | 10 | 12 | 9 | 12 | 14 | 9 | 14 | 17 | 25 |
| Pharmacology Pharmacy | | 4 | 2 | 13 | 5 | 8 | 7 | 10 | 7 | 7 | 17 |
| Biotechnology Applied Microbiology | 3 | 8 | 6 | 6 | 13 | 11 | 12 | 16 | 9 | 18 | 19 |
| Instruments Instrumentation | 12 | 5 | 16 | 16 | 11 | 10 | 9 | 17 | 9 | 12 | 15 |
| Microbiology | 1 | 5 | 3 | 10 | 7 | 11 | 14 | 12 | 11 | 14 | 17 |
| General & Internal Medicine | 4 | | 1 | 2 | 2 | 1 | 5 | 5 | 3 | 7 | 9 |

# Appendix C

## Yearly disciplinary breakdown of Indo-German collaborative research from 2012-2022

| Research area | 2012 | 2013 | 2014 | 2015 | 2016 | 2017 | 2018 | 2019 | 2020 | 2021 | 2022 |
|---|---|---|---|---|---|---|---|---|---|---|---|
| Chemistry | 165 | 182 | 250 | 221 | 265 | 298 | 284 | 307 | 287 | 352 | 296 |
| Physics | 208 | 165 | 209 | 176 | 183 | 147 | 167 | 228 | 220 | 200 | 150 |
| Astronomy & Astrophysics | 199 | 180 | 193 | 164 | 203 | 154 | 213 | 246 | 163 | 316 | 271 |
| Materials Science | 75 | 75 | 82 | 104 | 94 | 120 | 108 | 142 | 144 | 140 | 145 |
| Engineering | 64 | 79 | 73 | 115 | 140 | 115 | 116 | 136 | 130 | 151 | 123 |
| Science & Technology - Other Topics | 54 | 79 | 71 | 93 | 113 | 126 | 127 | 135 | 160 | 177 | 154 |
| Biochemistry & Molecular Biology | 60 | 59 | 74 | 82 | 65 | 61 | 77 | 105 | 104 | 139 | 112 |
| Computer Science | 43 | 50 | 69 | 80 | 117 | 103 | 102 | 131 | 126 | 144 | 135 |
| Environmental Sciences & Ecology | 45 | 30 | 55 | 52 | 52 | 52 | 77 | 74 | 111 | 128 | 143 |
| Mathematics | 22 | 20 | 19 | 32 | 37 | 42 | 44 | 53 | 44 | 65 | 70 |
| Crystallography | 9 | 3 | 2 | 12 | 4 | 9 | 14 | 5 | 10 | 10 | 5 |
| Oncology | 27 | 19 | 21 | 25 | 35 | 44 | 51 | 46 | 68 | 102 | 65 |
| Neurosciences & Neurology | 17 | 33 | 31 | 24 | 39 | 28 | 26 | 47 | 76 | 79 | 61 |
| Geology | 19 | 27 | 32 | 20 | 19 | 30 | 50 | 26 | 32 | 36 | 42 |
| Plant Sciences | 17 | 19 | 18 | 29 | 28 | 20 | 39 | 30 | 35 | 47 | 36 |
| Agriculture | 17 | 14 | 19 | 20 | 23 | 26 | 18 | 21 | 36 | 38 | 30 |
| Optics | 19 | 25 | 16 | 17 | 13 | 26 | 27 | 23 | 29 | 33 | 12 |

| | | | | | | | | | | |
|---|---|---|---|---|---|---|---|---|---|---|
| **Business & Economics** | 10 | 12 | 22 | 34 | 70 | 33 | 40 | 34 | 36 | 44 | 33 |
| **Geochemistry & Geophysics** | 17 | 14 | 16 | 23 | 18 | 20 | 15 | 23 | 16 | 24 | 32 |
| **Pharmacology & Pharmacy** | 11 | 18 | 30 | 24 | 19 | 23 | 28 | 25 | 39 | 43 | 26 |
| **Biotechnology & Applied Microbiology** | 10 | 15 | 18 | 20 | 15 | 27 | 21 | 22 | 31 | 25 | 22 |
| **Instruments & Instrumentation** | 21 | 14 | 23 | 11 | 24 | 17 | 12 | 19 | 19 | 19 | 9 |
| **Microbiology** | 20 | 19 | 18 | 13 | 14 | 13 | 15 | 15 | 24 | 24 | 38 |
| **General & Internal Medicine** | 3 | 13 | 7 | 11 | 10 | 36 | 27 | 43 | 37 | 52 | 49 |

# Appendix D

## Cluster analysis of Indo-German research publications based on co-authorship

| Cluster/No. of items | Organization | Country |
|---|---|---|
| **Cluster-1/10** | Inst High Energy Phys | China |
| | Panjab Univ | India |
| | Univ Sci & Technol China | China |
| | Univ Illinois | USA |
| | Charles Univ Prague | Czech Republic |
| | Brookhaven Natl Lab | USA |
| | Joint Inst Nucl Res | RUSSIA |
| | Indiana Univ | USA |
| | Korea Univ | Korea |
| | Univ Delhi | India |
| **Cluster-2/9** | Tata Inst Fundamental Res | India |
| | Johannes Gutenberg Univ Mainz | Germany |
| | Ist Nazl Fis Nucl | Italy |
| | Univ Manchester | England |
| | Rhein Westfal Th Aachen | Germany |
| | Univ Michigan | USA |
| | Princeton Univ | USA |
| | Univ Calif Berkeley | USA |
| | Caltech | USA |
| **Cluster-3/9** | Indian Inst Technol | India |

| | Tech Univ Munich | Germany |
| | Indian Inst Sci | India |
| | Univ Bonn | Germany |
| | Univ Tokyo | Japan |
| | Bhabha Atom Res Ctr | India |
| | Heidelberg Univ | Germany |
| | Chinese Acad Sci | China |
| | Tech Univ Darmstadt | Germany |